\documentclass[aps,prl,reprint,superscriptaddress]{revtex4-1}
\usepackage{graphicx,xcolor,soul}%
\usepackage{amsmath}%
\usepackage{amsfonts}%
\usepackage{amssymb}%
\usepackage{graphicx}
\usepackage{mathptmx}
\usepackage{CJK}
\usepackage[colorlinks=true,citecolor=blue,breaklinks=true,%
urlcolor=blue,linkcolor=blue]{hyperref}
\providecommand{\U}[1]{\protect\rule{.1in}{.1in}}
\def\unit#1{\mathord{\thinspace\rm #1}}

\usepackage[normalem]{ulem}
\usepackage{color}

\begin{document}
\begin{CJK*}{Bg5}{bsmi}
\title{Fabry-P\'{e}rot interference in gapped bilayer graphene with broken anti-Klein tunneling}
\author{Anastasia Varlet}
\email{varleta@phys.ethz.ch}
\affiliation{Solid State Physics Laboratory, ETH Z\"{u}rich, 8093 Z\"{u}rich, Switzerland}
\author{Ming-Hao Liu (¼B©ú»¨)}
\affiliation{Institut f\"{u}r Theoretische Physik, Universit\"{a}t Regensburg, D-93040
Regensburg, Germany}
\author{Viktor Krueckl}
\affiliation{Institut f\"{u}r Theoretische Physik, Universit\"{a}t Regensburg, D-93040
Regensburg, Germany}
\author{Dominik Bischoff}
\affiliation{Solid State Physics Laboratory, ETH Z\"{u}rich, 8093 Z\"{u}rich, Switzerland}
\author{Pauline Simonet}
\affiliation{Solid State Physics Laboratory, ETH Z\"{u}rich, 8093 Z\"{u}rich, Switzerland}
\author{Kenji Watanabe}
\affiliation{Advanced Materials Laboratory, National Institute for Materials Science, 1-1
Namiki, Tsukuba 305-0044, Japan}
\author{Takashi Taniguchi}
\affiliation{Advanced Materials Laboratory, National Institute for Materials Science, 1-1
Namiki, Tsukuba 305-0044, Japan}
\author{Klaus Richter}
\affiliation{Institut f\"{u}r Theoretische Physik, Universit\"{a}t Regensburg, D-93040
Regensburg, Germany}
\author{Klaus Ensslin}
\affiliation{Solid State Physics Laboratory, ETH Z\"{u}rich, 8093 Z\"{u}rich, Switzerland}
\author{Thomas Ihn}
\affiliation{Solid State Physics Laboratory, ETH Z\"{u}rich, 8093 Z\"{u}rich, Switzerland}

\date{\today}

\begin{abstract}
We report the experimental observation of Fabry-P\'{e}rot (FP) interference in the conductance of a gate-defined cavity in a dual-gated bilayer graphene (BLG) device. The high quality of the BLG flake, combined with the device's electrical robustness provided by the encapsulation between two hexagonal boron nitride layers, allows us to observe ballistic phase-coherent transport through a $1\unit{\mu m}$-long cavity. We confirm the origin of the observed interference pattern by comparing to tight-binding calculations accounting for the gate-tunable bandgap. The good agreement between experiment and theory, free of tuning parameters, further verifies that a gap opens in our device. The gap is shown to destroy the perfect reflection for electrons traversing the barrier with normal incidence (anti-Klein tunneling). The broken anti-Klein tunneling implies that the Berry phase, which is found to vary with the gate voltages, is always involved in the FP oscillations regardless of the magnetic field, in sharp contrast with single-layer graphene.

\end{abstract}

\maketitle

\end{CJK*}

Interference of particles is a manifestation of the wave nature of matter. A well-known realization is the double-slit experiment, which cannot be described by the laws of Newtonian mechanics, but requires a full quantum description. This experiment has been performed with photons \cite{Taylor1909,Aspect1986}, electrons \cite{Mollenstedt_1956} and even molecules \cite{arndt_waveparticle_1999}. Another setting widely used in optics is the Fabry-P\'{e}rot (FP) interferometer, where a photon bounces back and forth between two coplanar semitransparent mirrors. Partial waves transmitted after a distinct number of reflections within this cavity interfere and give rise to an oscillatory intensity of the transmitted beam as the mirror separation or the particle energy is varied.

In solid-state physics, graphene has proven to be a suitable material for probing electron interference at cryogenic temperatures \cite{Russo_AB_2008,Huefner_2010}. However, in single-layer graphene (SLG) the realization of FP interferometers is challenging. The absence of a bandgap and the Klein tunneling hamper the efficiency of sharp potential steps between n- and p-type regions, which play the role of the interferometer mirrors \cite{Katsnelson2006,Shytov2009,Tudorovskiy2012}. Theory suggests that smooth barriers enhance the visibility of interference \cite{Young2009,Velasco2009} due to Klein collimation \cite{Cheianov2006}. Recently, ultraclean suspended SLG devices have shown FP interference with stunning contrast using cavity sizes of more than $1\unit{\mu m}$ \cite{Rickhaus2013,Grushina2013,Hakonen2014}.

In bilayer graphene (BLG) potential steps between n- and p-type regions lead to evanescent interface states resulting in a zero-transmission at normal incidence, known as anti-Klein tunneling~\cite{Katsnelson2006}. Furthermore, in BLG a bandgap can be induced by a transverse electric field \cite{mccann_asymmetry_2006,ohta_controlling_2006,castro_biased_2007,oostinga_gate-induced_2008}. The Berry phase of $\pi$ in SLG has been predicted \cite{Shytov2008} and observed \cite{Young2009} to cause a phase jump of $\pi$ in the FP fringes at weak magnetic field ($B$). In gapless BLG the Berry phase is known to be $2\pi$, but it is yet to be understood how the Berry phase in gapped BLG influences the FP interference.

In this paper, we investigate the FP interference in gapped BLG. Comparing experiment and theory, we reveal the roles played by the anti-Klein tunneling, the gap and the Berry phase. We perform transport measurements on a locally-dual-gated BLG encapsulated between hexagonal boron nitride (h-BN) flakes. The high quality of the dual-gated area allows us to observe FP interference in the bipolar regime. The origin of the measured oscillations is confirmed by ballistic transport simulations. The dependencies of the interference patterns on the carrier density, the $B$-field, the applied bias, and the temperature are investigated. We consistently find signatures of ballistic phase-coherent transport over a $1~\rm{\mu m}$ distance and highlight the very different tunneling behavior of BLG and SLG.

\begin{figure}[t]
\includegraphics[width=\columnwidth]{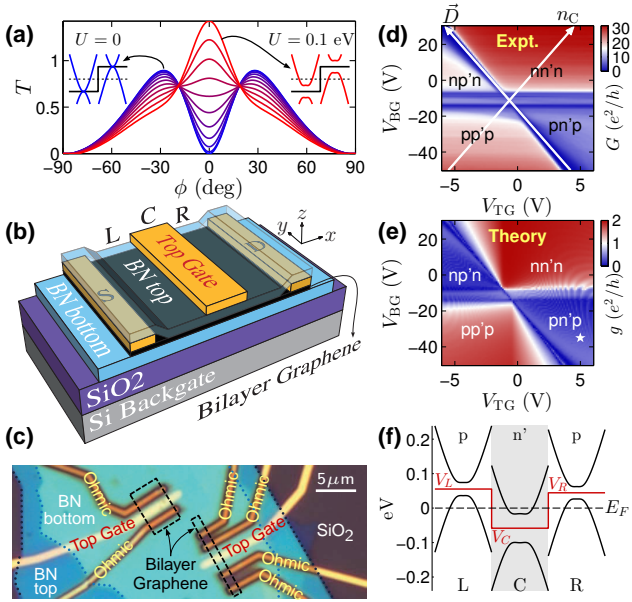}
\caption{(color online). (a) Transmission $T$ as a function of the incidence angle $\phi$ for an electron traversing a symmetric, sharp np junction, with changing asymmetry parameters $U=0,0.01,\cdots,0.1\unit{eV}$. (b) Schematics of the device: a BLG flake is sandwiched between two h-BN layers. (c) Optical microscope image of two devices fabricated from the same flakes. The measurements were carried out on the right device. (d) Conductance map measured at $1.6\unit{K}$. (e) Normalized conductance map calculated based on the model depicted in (f) with the configuration marked by the star in (e).}
\label{fig1}
\end{figure}

We first point out a simple but striking theoretical fact regarding a single junction. The perfect reflection across a bipolar potential step in BLG was predicted in Ref.\ \onlinecite{Katsnelson2006}, in the absence of the gap. Little attention has been paid, however, to the fact that the presence of the gap destroys the perfect reflection at normal incidence \cite{Park2011,Rudner_2011}. By calculating the angle-resolved transmission $T(\phi)$  (Green's function method based on a tight-binding model as in Ref.\ \onlinecite{Liu2012}), we find that a non-zero normal incidence transmission can be recovered by introducing a gap. As illustrated in Fig.\ \ref{fig1}(a), $T(\phi=0)$ increases with the so-called asymmetry parameter $U$ defined as the on-site energy difference between the two graphene layers \cite{McCann2013}. Here we consider transport through an ideal, symmetric, sharp n-p junction with the potential step height fixed at $0.2\unit{eV}$ and $U$ increasing from zero to $0.1\unit{eV}$. Due to the two-fold valley degeneracy incorporated in the tight-binding model, the transmission function $T$ has a maximum possible value of $2$ within the single-band regime (i.e., energy $|E|\leq \gamma_1$, $\gamma_1\approx 0.39\unit{eV}$ being the nearest-neighbor interlayer hopping). As seen in Fig.\ \ref{fig1}(a), $T(0)=0$ only when $U=0$, which is the celebrated anti-Klein tunneling. It acquires finite values when $U\neq 0$. Recovering a finite $T(0)$ with increasing bandgap thus reveals the counter-intuitive role played by $U$, which increases the transmission by suppressing anti-Klein tunneling.

The investigated device, sketched in Fig.\ \ref{fig1}(b) and (c) (right device), consists of a h-BN/BLG/h-BN stack deposited on a Si/SiO$_{2}$-substrate, prepared as presented in Ref.\ \onlinecite{Varlet2014} using the transfer technique described in Ref.\ \onlinecite{Dean2010}. The measurements were performed at a temperature of $1.6\unit{K}$ in a two-terminal configuration (inner ohmic contacts). Unless stated otherwise, a constant bias voltage was applied symmetrically between the ohmic contacts [S and D in Fig.\ \ref{fig1}(b)] and the current was measured. Modulating the top-gate voltage with an ac component enabled the measurement of the normalized transconductance $dG/dV_{\rm TG}$.

Our device consists of three areas: two outer areas (denoted as L and R, of length $\ell_{\rm L}=\ell_{\rm R}=0.95\unit{\mu m}$) simultaneously tuned by the voltage $V_\mathrm{BG}$, and the central area (denoted as C, $\ell_{\rm C}=1.1\unit{\mu m}$) which is tuned by both top- and back-gate ($V_\mathrm{TG}$,$V_\mathrm{BG}$). The two-dimensional map presented in Fig.\ \ref{fig1}(d) shows the conductance of the device measured as a function of both $V_\mathrm{TG}$ and $V_\mathrm{BG}$. Depending on the applied voltages, four different polarity combinations are possible: two with same polarities (pp'p and nn'n - the prime referring to region C) and two with opposite ones (np'n and pn'p). The charge neutrality point of the dual-gated area evolves along the so-called displacement field $\vec{D}$ axis. While increasing $|\vec{D}|$, the conductance at the charge neutrality point decreases because a gap is opened. This bandgap has been characterized with temperature-dependent measurements, which provided the activated gap size of $E_{\rm gap} = 2\unit{meV}$ at $V_\mathrm{BG} = 28\unit{V}$ (corresponding to $\vec{D} = 1.1\unit{V/nm}$). As previously reported \cite{Oostinga2008,Russo2009,Weitz2010,Thiti_2010}, transport in the gap is dominated by hopping, preventing the experimental determination of the real gap size.

Our calculations for the gate-dependent conductance use a Green's function formalism~\cite{Rickhaus2013,Liu2012a} based on the
nearest-neighbor tight-binding model for BLG~\cite{Slonczewski1958,McClure1957} with the on-site energy profile extracted from the gate-modulated carrier density. To relate the carrier density to the gate voltages in the respective areas, we adopt a parallel-plate capacitance model \footnote{See supplemental material for details of the electrostatic model, gate-dependence of the asymmetry parameter and the Berry phase for the device, as well as more supporting experimental data. It includes Ref.~\cite{Berry1984}}. In area $X$ ($X=\mathrm{L,C,R}$), the carrier density $n_X=n_X(V_{\rm TG},V_{\rm BG})$ is composed of top-gate contribution $n_X^t$ (only for $X=\mathrm{C}$), back-gate contribution $n_X^b$, and intrinsic doping contribution $n_X^0$. Following the tight-binding theory of the gap~\cite{McCann2013}, the inputs of $n_X^t,n_X^b,n_X^0$ allow us to compute the asymmetry parameters $U_X(V_{\rm TG},V_{\rm BG})$ \cite{Note1}. Finally, with $n_X$ and $U_X$, the band offset for region $X$ to be added to the diagonal matrix elements (on-site energy) of the tight-binding Hamiltonian $V_X$ is calculated (see \cite{Note1} for the expression of $V_X$): this fixes the global Fermi level at energy $E=0$ where linear-response transport occurs.

\begin{figure}[b]
\includegraphics[width=\columnwidth]{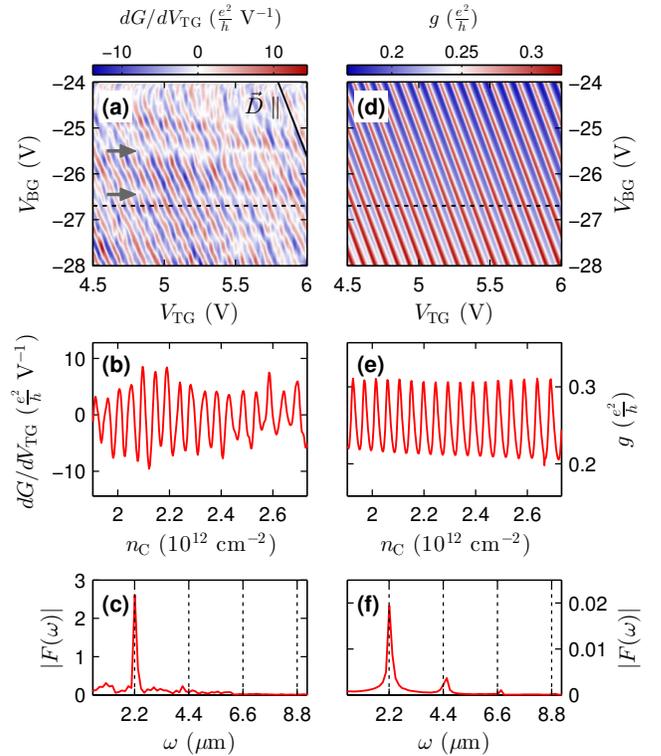}
\caption{(color online). (a) Measured transconductance and (d) calculated normalized conductance in the pn'p region. In (a), the grey arrows indicate faint horizontal patterns which originate from the L and R regions and the slope of the $\vec{D}$-axis is shown with a black line. (b)/(e) Cut along the dashed line in (a)/(d). By projecting from the density $n_{\rm C}$ onto the wave-vector $k$ axis, the discrete Fourier transforms $F(\omega)$ of (b) and (e) are shown in (c) and (f), respectively, both of which exhibit a sharp peak at frequency $\omega=2.2\unit{\mu m}$, which is precisely twice the cavity length $2\ell_{\rm C}$.}
\label{fig2}
\end{figure}

\begin{figure}[t]
\includegraphics[width=\columnwidth]{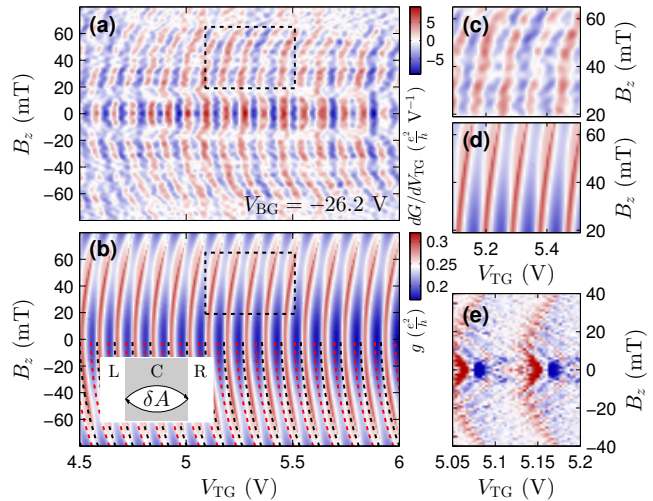}
\caption{(color online). (a) Measured and (b) calculated $B$-field dependence of the FP oscillations at $V_{\rm BG} = -26.2~\rm{V}$. Inset of (b) shows a closed loop considered in Eq.\ \eqref{res cond}. Black and red dashed curves (sketched only for $B_z\leq 0$ for clarity) are numerical solutions to the resonance condition \eqref{res cond} with and without the Berry phase, respectively. (c) and (d) are zoom-in from (a) and (b), respectively, indicated by the black dashed boxes. (e) Calculated $dg/dV_{\rm TG}$ map for a model with the scattering region simulating the full $\mathrm{L,C,R}$ areas.}
\label{fig3}
\end{figure}

As an example illustrating the model, the band diagram at gate voltages $(V_{\rm TG},V_{\rm BG})=(5,-35)\unit{V}$ is sketched in Fig.\ \ref{fig1}(f), where the band offset level $V_{X}$ takes as inputs $n_X$ and $U_X$. To keep the model as simple as possible, we neglect stray fields of the top-gate and assume the potential profile to be ideally flat and sharp. We focus on the transport property of the dual-gated region, modeling L and R as semi-infinite leads and regarding C as the scattering region [with the only exception in Fig.\ \ref{fig3}(e); see below]. Such a simplified model taking into account the ideal bandgap \cite{McCann2013} turns out to work surprisingly well, as the main features of the measured conductance map [Fig.\ \ref{fig1}(d)] are well-captured by our calculation shown in Fig.\ \ref{fig1}(e). Note that different from Ref.\ \onlinecite{Rickhaus2013}, throughout this work we will not worry about mode counting and concentrate only on the normalized conductance $g$ (with maximum of 2 due to valley degeneracy).

We now focus on the pn'p regime: figure \ref{fig2}(a) depicts the normalized transconductance, showing a pronounced oscillatory behavior (already visible in the conductance \cite{Note1}). These oscillations evolve parallel to the $\vec{D}$-axis [for comparison, the slope of the $\vec{D}$-axis is displayed in Fig.\ \ref{fig2}(a) as a black line], indicating that they are related to the dual-gated part. In Fig.\ \ref{fig2}(b), an example of a transconductance trace taken at $V_\mathrm{BG} = -26.7\unit{V}$ is shown. To confirm the origin of the oscillatory signal, we analyze the oscillation frequency of Fig.\ \ref{fig2}(b) by first projecting the transconductance from the displayed density axis, $n_{\rm C}$, onto the wave-vector axis, $k = \sqrt{\pi|n_{\rm C}|}$, and then performing the discrete Fourier transform from $k$ to frequency $\omega$ space. Since the main contribution to the FP interference is given by phase difference $\Delta\Phi = k\cdot 2\ell_{\rm C}$, the oscillation frequency is expected to be $2\ell_{\rm C} = 2.2\unit{\mu m}$, which is in agreement with our observations in Fig.\ \ref{fig2}(c). Similar behavior has been observed for different density ranges in the pn'p regime, indicating the robustness of the phenomenon \cite{Note1}.

Figures \ref{fig2}(d--f) are the theory counter parts of experimental Figs.\ \ref{fig2}(a--c), except that the theoretical analysis is done from the normalized conductance (the calculation is ideally ballistic and therefore leads to a clear-enough visibility). We emphasize that the calculation was done based on the simplified model illustrated in Fig.\ \ref{fig1}(f), free of any tuning parameters. In Fig.\ \ref{fig2}(f), additionally to the main peak, higher harmonics are visible at integer multiples of the fundamental period. The agreement between our theory and experiment is satisfactory, confirming the ballistic origin of the fringes reported in Fig.\ \ref{fig2}(a) as FP interference in the dual-gated area C.

To further understand the FP interference in BLG, we performed magneto-transport measurements and calculations in the low $B$-field regime, where the feature of the conductance is still dominated by FP interference without entering the quantum Hall regime. Figure \ref{fig3} shows the measured gate-field map of $dG/dV_{\rm TG}$ (a) and the calculated one of $g$ (b). Despite some differences in the $B$-field dependence of the FP fringes in the two maps for the field range of $|B_z|\lesssim 20\unit{mT}$, (i) the number of oscillation periods and (ii) the field dependence for the range $20\unit{mT}\lesssim |B_z|\lesssim 60\unit{mT}$, are consistent. The former can be directly counted (both 17 stripes), while the latter can be more clearly seen by comparing Figs.\ \ref{fig3}(c) and \ref{fig3}(d) that are zoom-in plots from Figs.\ \ref{fig3}(a) and \ref{fig3}(b), respectively, indicated by the dashed boxes therein.

To elucidate the peculiar field behavior observed only in the experimental
map of Fig.\ \ref{fig3}(a) at $|B_z|\lesssim 20\unit{mT}$, we investigate
the accumulated phases for a closed loop  in the dual-gated cavity $\mathrm{C}$
given by a classical trajectory encircling, arising from backscattering at the interfaces, the area $\delta A$,
as sketched in the inset of Fig.\ \ref{fig3}(b).
The total phase can be decomposed into a kinetic part $\Phi_{\rm WKB}$,
the Aharonov-Bohm phase $\Phi_{\rm AB}=eB_z\delta A/\hbar$
and the Berry phase $\Phi_\mathrm{Berry}$.
Based on the phase difference between a transmitted and a twice reflected electron wave, we get the
FP resonance condition
\begin{equation}\label{res cond}
\Phi_{\rm WKB}+\Phi_{\rm AB}+\Phi_{\rm Berry}=2{\pi}j,\quad j\in\mathbb{Z},
\end{equation}
where the $B$-field dependence only enters through the first two terms.
As shown in more detail in \cite{Note1}, our computation of the Berry phase of BLG shows that it is only $2\pi$ for vanishing asymmetry $U_{\rm C}$ and can generally take values between $0$ and $2\pi$ depending on the carrier density $n_{\rm C}$ and layer asymmetry $U_{\rm C}$. For the bipolar region presented in Fig.~\ref{fig2}(a), $\Phi_\mathrm{Berry}$ is nearly constant, ranging between $1.22 \pi$ and $1.46 \pi$~\cite{Note1}.
 By numerically solving Eq.\ \eqref{res cond}, we find a set of resonance contours well matching the periodicity of the FP fringes and being located quite closely to the conductance maxima [see black dashed contours in Fig.\ \ref{fig3}(b)]. Furthermore, neglecting $\Phi_{\rm Berry}$ in Eq.\ \eqref{res cond} leads to the red dashed contours, which do not coincide with the positions of constructive interference. We therefore conclude that an effect of the Berry phase exists in the transport calculation, and is always involved independent of $B_z$. This is in sharp contrast to the case of SLG, where the Berry phase effect requires a weak $B$-field in order to overcome the perfect transmission of Klein tunneling \cite{Shytov2008}. The experimentally observed phase jump of $\pi$ in SLG \cite{Young2009} was previously reproduced by two of us \cite{Liu2012a}
based on the same method.

From the above discussion, the Berry phase can be ruled out as a possible reason for the peculiar $B$-field behavior observed for $|B_z|\lesssim 20\unit{mT}$ in the experiment in Fig.\ \ref{fig3}(a). We next suspect that the effect originates from the outer areas L and R. While the fringes reported in Fig.\ \ref{fig2}(a) have been confirmed to arise from FP interference in the C area, the L and R areas that are independent of $V_{\rm TG}$ and slightly shorter than C may exhibit FP interference as well. Indeed by a closer look at Fig.\ \ref{fig2}(a) one can identify a few horizontal patterns (see grey arrows) that are independent of $V_{\rm TG}$ and may be attributed to FP interference within the outer regions L and R. Taking the cavity length of $\ell_{\rm L}=\ell_{\rm R}=0.95\unit{\mu m}$, we estimate the expected voltage spacing to be around $0.6\unit{V}$ consistent with those horizontal patterns observable in Fig.\ \ref{fig2}(a). By performing a ballistic calculation taking into account the total $3\unit{\mu m}$ length of the $\mathrm{L,C,R}$ areas, we show in Fig.\ \ref{fig3}(e) a gate-field map focusing on a smaller $V_{\rm TG}$ range. The peculiar behavior at low field close to zero is indeed recovered, supporting the idea that the outer areas are the main cause. The interplay between the different field dependencies of the FP interference in the $\mathrm{L,C,R}$ cavities is, however, beyond the scope of the present discussion.

\begin{figure}[t]
\includegraphics[width=\columnwidth]{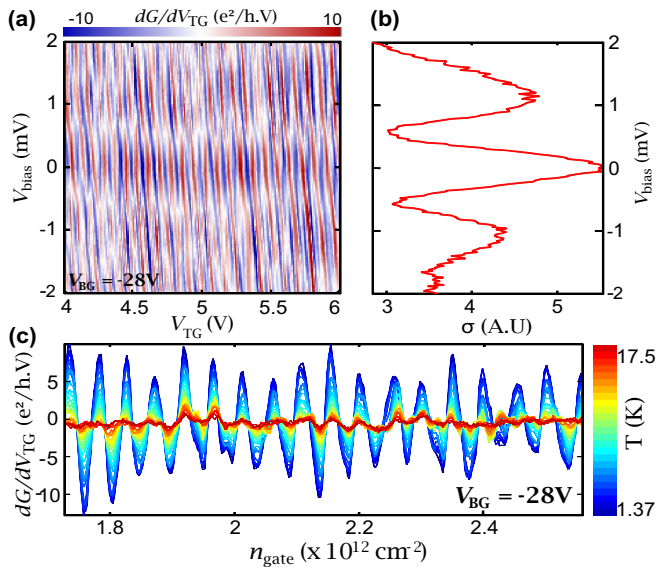}
\caption{(color online). (a) Bias dependence of the FP oscillations. (b) Standard deviation of the oscillations as a function of the bias. (c) Temperature dependence of the oscillations.}
\label{fig4}
\end{figure}

We now focus on the experimental bias dependence of the FP-modulated signal shown in Fig.\ \ref{fig4}(a). The oscillations undergo a linear shift as a function of the source-drain bias. On top of this, their amplitude is modulated. To highlight this modulation, we calculate, at each bias value, the standard deviation from the mean of the signal [Fig.\ \ref{fig4}(b)]. A modulation is seen, with two minima at $\sim \pm 0.6\unit{mV}$. This dependence reflects the energy averaging effect of the bias voltage. We may understand the situation by considering the oscillating transmission as a function of energy within the bias window $eV_{\rm bias}$: each time the window encloses an integer number of oscillation periods, a minimum is reached in the signal.

Finally, we analyze the experimental effect of temperature $T$ on the FP interference in Fig.\ \ref{fig4}(c) with the range $1.37\unit{K}\leq T \leq 17.5\unit{K}$. We can distinguish two regimes: at low-$T$, the effect of the temperature on the oscillation amplitude is strong; however, at higher $T$, the effect seems to saturate with some persistent oscillations. These almost temperature-independent features are not of coherent origin and we therefore subtract their contribution from the oscillations for the following analysis. We then convolute the lowest temperature curve with the derivative of the Fermi-Dirac distribution for the whole range of available temperatures to understand the behavior of the amplitude of the oscillations as a function of $T$ \cite{Note1}. We find that the behavior is well described by thermal damping of the oscillations: $dG/dV_{\rm TG} \sim A  \xi {\text{csch}({\xi})}$, with $A$ being a scaling parameter, $\xi = 2\pi^2 k_{b}T/\Delta E$ and $\Delta E = 1.2\unit{meV}$ being the averaged spacing of the oscillations in the studied interval.

In conclusion, we observed Fabry-P\'{e}rot oscillations in a $1\unit{\mu m}$-long gapped BLG cavity. We characterized the origin of these oscillations studying their density, $B$-field, bias and temperature dependencies. Our calculations were able to reproduce our observation and therefore demonstrated the importance of the tunable bandgap, which leads to a lifting of anti-Klein tunneling. This allowed us to confirm the ballistic phase-coherent nature of transport through the dual-gated region. Our work combined with the recent advances in the quality of sandwiched structures \cite{Wang2013} is a step towards future electron optics experiments in gapped BLG.

We thank Aleksey Kozikov, Fabrizio Nichele and Stephan Baer for constructive comments and fruitful discussions. We also acknowledge financial support from the Marie Curie ITNs $S^3$NANO and QNET, together with the Swiss National Science Foundation via NCCR Quantum Science and Technology, the Graphene Flagship and the Deutsche Forschungsgemeinschaft within SFB $689$ and SPP $1666$.

\bibliographystyle{apsrev4-1}
\bibliography{FPpapers}

\end{document}